\documentclass[fleqn,10pt]{wlscirep}
\usepackage[utf8]{inputenc}
\usepackage[T1]{fontenc}
\usepackage{gensymb}
\usepackage{tabularx}
\usepackage{multicol,array}
\usepackage{subcaption}
\usepackage{tikz}
\usepackage{balance}
\usepackage[font=small,justification=justified,format=plain]{caption}
\usepackage{titlesec}

\newenvironment{Figure}
  {\par\medskip\noindent\minipage{\linewidth}}
  {\endminipage\par\medskip}

\makeatletter
\renewcommand{\fnum@figure}{Fig. \thefigure}
\makeatother

\newcolumntype{C}[1]{>{\centering\arraybackslash}p{#1}}

\title{Dual-comb spectroscopy for the characterization of laboratory flames}

\author[1]{Bernat Frangi}
\author[1]{Laura Monroy}
\author[1]{Aldo Moreno-Oyervides}
\author[1]{Oscar El\'ias Bonilla-Manrique}
\author[2]{Mariano Rubio-Rubio}
\author[2]{Mario S\'anchez-Sanz}
\author[1]{Pedro Mart\'in-Mateos}

\affil[1]{Department of Electronic Technology, Universidad Carlos III de Madrid, 28911 Legan\'es, Spain}
\affil[2]{Department of Thermal and Fluids Engineering, Universidad Carlos III de Madrid, 28911 Legan\'es, Spain}
\affil[*]{Corresponding author: bernat.frangi@uc3m.es}

\begin{abstract}

Optical spectroscopy, in particular dual-comb (DC) spectroscopy, is a critical, non-invasive tool for combustion diagnostics, offering high precision and calibration-free advantages. However, its implementation remains challenging, especially in the mid-infrared region. This work presents the development of a robust DC spectroscopic system based on electro-optical (EO) frequency comb generators and difference frequency generation (DFG), specifically designed for the characterization of laboratory flames. Operating at a center wavelength of 3427.43\,nm, the system utilizes a differential detection strategy to enable precise, calibration-free measurements of unburned methane ($\mathrm{CH_{4}}$) concentrations in a McKenna burner. The experimental results demonstrate an estimated detection limit of 1.1\,ppm for a 1\,m path length and effectively resolve spatial concentration gradients across the combustion region. Furthermore, the system's high temporal resolution allowed for the identification of dynamic combustion instabilities, including self-sustained pulsations and fuel leakage under fuel-lean conditions. These findings validate the proposed EO architecture as a flexible and highly sensitive tool for advanced flame characterization.\\

\end{abstract}

\begin{document}

\flushbottom
\maketitle
\thispagestyle{empty}

\begin{multicols}{2}

\section{Introduction}

Optical metrology and spectroscopy are important and increasingly standardized tools in many fields, ranging from astronomy to time measurement and the analysis of gaseous samples. One such field is the analysis of combustion flames in laboratories for the development and refinement of numerical models \cite{CURRAN201957}, an area in which optical spectroscopy has shown enormous potential, with new developments continuing to emerge \cite{Kohse-Hoinghaus2023}. Compared to traditional invasive methods, such as the use of thermocouples, optical methods promise to provide accurate, remote measurements of parameters of interest in flames without affecting the combustion process \cite{webber2000situ, ZHANG2024109344}.

In this regard, a large number of experimental demonstrations have recently been carried out that take advantage of optical absorption and dispersion measurements. Among the absorption-based methods employed, Wavelength Modulation Spectroscopy (WMS) \cite{Chao2012} stands out for having been used in a large number of studies. However, the emergence of widely tunable Quantum Cascade Lasers (QCL) has led to the growing importance of measurement approaches based on tunable diode laser absorption spectroscopy (TDLAS) scanning \cite{ZHANG2024109344}. Another method that is growing in popularity is dispersion spectroscopy \cite{photonics12060537, Ma2018, 10494328}. Unlike absorption spectroscopy, it extracts information from the refractive index profile rather than absorption and it has been used in numerous combustion studies in recent years. Furthermore, when discussing high-performance optical spectroscopy today, studies conducted using optical frequency combs (specifically dual-comb, or DC) \cite{Coddington:16} are hard to avoid, and the field of combustion research is no exception, with some notable work having been presented \cite{Yun:23, Takeshi:25, Xu10092022, SCHROEDER20174565, COBURN2024105533}. Compared to virtually any other spectroscopic measurement method, the use of frequency combs offers much higher frequency accuracy in determining the shape of the absorption line, high measurement speeds, and has a fundamental advantage in that no calibration is required. This significantly impacts data processing, as spectral characterizations can be directly introduced into models, eliminating the need for additional experimental characterizations or complex baseline compensation/elimination algorithms. This results in more accurate and straightforward extraction of the desired information from the spectral measurement. In addition, the simultaneous acquisition of all spectral components makes DCS inherently robust to beam steering, preserving the absorption line shape and avoiding the need for complex correction algorithms.

Despite their enormous capabilities, implementing and using DC systems remains challenging today. Early demonstrations have focused on the near-infrared region, where frequency comb oscillators, such as mode-locked fiber lasers, are more readily available, highlighting the exceptional spectral resolution achievable with DC spectroscopy \cite{PhysRevLett.100.013902, Zolot:12}. However, because the absorption features accessible in the near-infrared correspond primarily to weak overtone transitions, obtaining measurable absorbance typically requires either long averaging times or extended effective optical path lengths, commonly realized using multipass cells. Moreover, the environmental sensitivity of mode-locked lasers, together with the very low power-per-tooth distributed across their inherently broad bandwidths, complicates their deployment in unsteady, single-pass combustion environments.

More recently, the emergence of mid-infrared QCLs and Kerr frequency combs has enabled numerous dual-comb implementations in the mid-infrared spectral region \cite{Hugi2012, Wang2013}. In this regime, most molecules of interest exhibit strong fundamental vibrational transitions, allowing substantially enhanced sensitivity and facilitating measurements under more representative combustion conditions. Nevertheless, limitations remain regarding the precise, flexible control of these comb sources. Kerr frequency combs typically generate repetition rates in the tens to hundreds of gigahertz, which are often too coarse to accurately resolve individual absorption lines \cite{PASQUAZI20181}. Similarly, although QCL frequency combs provide high optical power in the mid-infrared, they can present challenges in simultaneously achieving robust mutual coherence and flexible tunability \cite{Consolino2019}. Consequently, there remains a clear need for an architecture that combines the intrinsic sensitivity of the mid-infrared with reduced stabilization complexity and highly agile, user-defined spectral sampling.

In this scenario, Electro-optical (EO) frequency comb generators \cite{Long:14} offer a comparatively simple and robust platform, having slowly but surely matured into remarkable technology, albeit at near-infrared wavelengths. This approach provides exceptional control over the comb teeth, supporting octave-spanning bandwidths \cite{Beha:17}, enabling direct dual-comb hyperspectral imaging \cite{Martin-Mateos:20}, and allowing robust, fast, and straightforward implementations for the analysis of one or a limited number of spectral features \cite{Millot2016}. In this regard, it is also worth highlighting the performance offered by cavity-enhanced DC implementations \cite{SUN2024105662}, which multiply the sensitivity of standard DC systems. Furthermore, by combining EO combs with nonlinear frequency conversion techniques such as difference-frequency generation (DFG), an effective pathway to mid-infrared DC spectroscopy is enabled. 

The main objective of this manuscript is to develop a DC optical spectroscopy system with EO generation combined with DFG, together with the configuration and data processing strategies necessary for the characterization of laboratory flames. The primary aim of this preliminary study was to analyze unburned fuel. To this end, the system was employed to examine various methane ($\mathrm{CH_{4}}$) flames in a McKenna burner. This is an application in which, as justified in the final sections of this manuscript, the proposed system has been able to obtain highly relevant information.

\section{System design and implementation}

The experimental setup is shown in \textbf{Fig. \ref{fig:setup1}}. The generation of the two combs was performed by electro-optically modulating a single continuous tunable laser source (Toptica, CTL1550) with a nominal spectral spanning of 130\,nm. This laser diode is driven in continuous-wave operation at 346\,mA and $25\degree$C, provided by an external controller (Toptica, DLC Pro) with low current noise and high temperature control for stable operation. The power and wavelength of the laser were set to $+16$\,dBm and 1541.05\,nm, as explained below. The output light is then injected into a dual-comb generation module. 

\begin{Figure}
    \centering
    \includegraphics[width=1\textwidth]{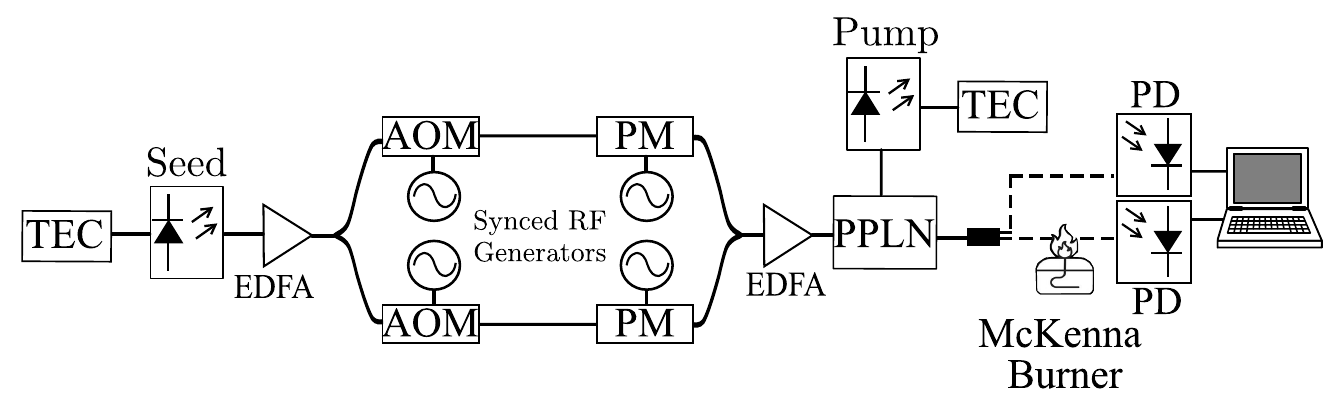}
    \captionof{figure}{Experimental setup for dual-comb generation and flame characterization. TEC: temperature controller, EDFA: erbium-doped fiber amplifier, AOM: acousto-optic modulator, PM: phase-modulator, PPLN: DFG module, PD: photodetector.}
    \label{fig:setup1}
\end{Figure}

In the first stage of this system, the incoming light is amplified by an erbium-doped fiber amplifier (EDFA) (Thorlabs, EDFA100S) to approximately 20\,dBm, which is then split equally into two optical branches using a $50:50$ fiber coupler. Each branch incorporates an acousto-optic modulator (AOM) (Gooch and Housego, T-M040-0.5C8J-3-F2S) to introduce a slight frequency shift ($f_\text{shift}$) ensuring a well-defined offset between the two branches \cite{martin2020}. A radio-frequency (RF) input voltage of 150\,mV, provided by a phase coherent Arbitrary Waveform Generator (AWG) (Rigol, DG4162) was adjusted for the driving signals of both AOMs generating a $f_\text{shift}$ of 40\,MHz and 40\,MHz $+$ 40\,kHz, respectively. Additionally, both branches include lithium niobate-based phase modulators (PM) (EOSpace) driven by a two-channel RF signal generator (Anapico, APMS20G-2), with a $V_\pi$ of 3.3\,V, enabling the efficient generation of frequency combs at low driving signals. The output combs are then amplified by a post-amplifier with 20\,dB, and modulated with slightly different modulation frequencies for each channel ($f_{r1}$ and $f_{r2}= f_{r1} +\Delta f$, where $\Delta f$ is the detuning between modulation frequencies), which set the spacing between the generated comb teeth. In this study, the system was operated with two different tooth spacing values to assess its performance under optimal conditions, as will be detailed in the next sections.

Once formed, the two combs are recombined via a second fiber coupler, producing a stable and tunable near-infrared region (NIR) dual-comb seed. A small fraction (1\%) of the output power is used to monitor the RF dual comb after photodetection in an InGaAs detector (Thorlabs, PDA10CF-EC). To preserve coherence over extended acquisition periods, particular care is taken to match optical path lengths and passively stabilize environmental conditions within the setup. The entire optical subsystem is enclosed in a rigid housing, and the temperature-sensitive components, especially the RF electronics are placed in thermally isolated compartments with adequate airflow. 

The dual-comb signal is subsequently amplified by an EDFA up to 125\,mW and frequency down-converted to the mid-infrared region (MIR) using a DFG process in a periodically poled lithium niobate (PPLN) crystal (NTT Electronics, WD-3440-000-A-B-C), achieving an efficiency of 13\%, which can be considered effectively constant across the entire comb bandwidth. The NIR combs are combined with a narrow-linewidth pump laser operating at 1064\,nm (Optilab, DFB-1064-200-CW), driven by a low noise current and temperature controller and delivering around 200\,mW of optical power. The two signals are coupled into the PPLN crystal through a wavelength division multiplexer (Opneti). Efficient DFG is ensured by precise thermal control of the nonlinear medium to maintain phase matching \cite{khan2023}. The resulting MIR output, centered at 3427.43\,nm, is collimated using a parabolic mirror and passed through a filter to remove residual NIR components.

For the final stage, the mid-infrared dual-comb output is guided through a dedicated mid-infrared fiber (Thorlabs, MZ22L1) towards a motorized XY translation platform (Standa, 8MT173-30) equipped with a beam collimator system, allowing the light to be precisely delivered and collimated into free space, facilitating long-distance measurements. Such an arrangement enables spatially resolved measurements of the sample by translating the measurement arm across the combustion region, as illustrated in \textbf{Fig. \ref{fig:setup2}}. Besides, a beam splitter is employed to divide the MIR dual comb into two separate free-space paths. One of these serves as a reference, while the second traverses the sample under test (SUT). Finally, the two beams are independently detected by two fast-response HgCdTe detectors (Vigo Photonics, PVI-4TE-8-1x1-TO8) optimized for mid-infrared detection. 

\begin{Figure}
    \centering
    \includegraphics[width=1\textwidth]{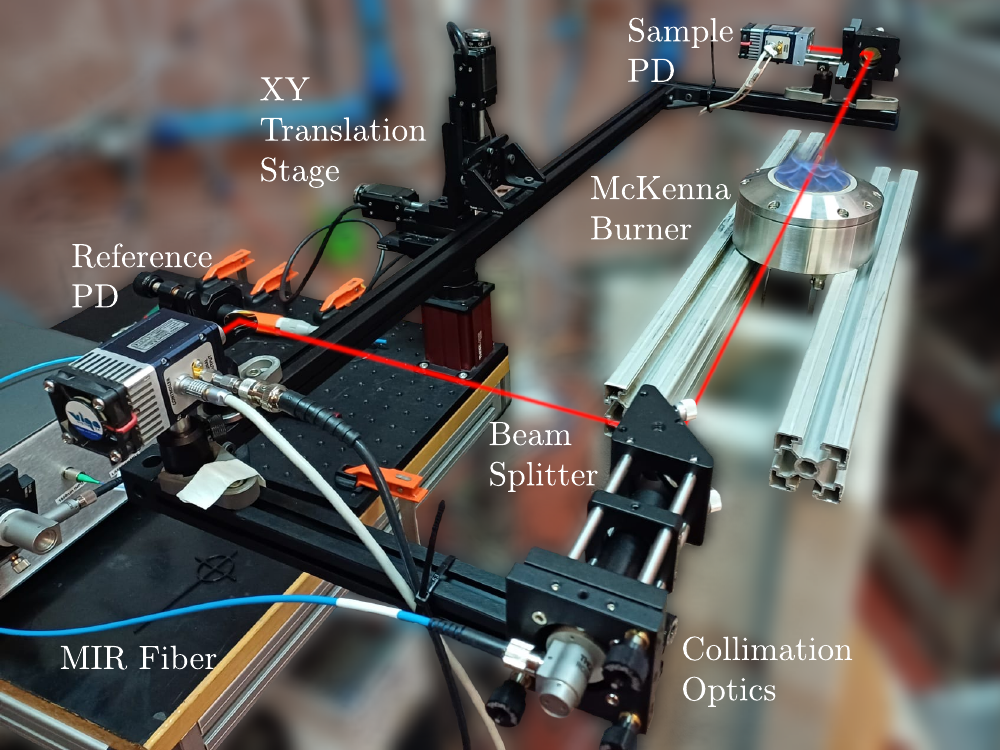}
    \captionof{figure}{Translation platform integrating the optical components and mid-infrared photodetectors for flame measurements. The laser propagation path is indicated for visualization purposes.}
    \label{fig:setup2}
\end{Figure}
The dual-channel detection strategy enables differential measurements by comparing the sample signal against the reference. This approach compensates for the power variations of the comb teeth, enabling a precise spectral measurement of the SUT without previous calibration. An advanced real-time processing algorithm is applied to the detector outputs, performing Fast Fourier Transforms (FFT) and normalization to retrieve the absorption spectrum of the SUT. This method allows precise reconstruction of spectral features, such as temperature and concentration, with high resolution and excellent signal-to-noise ratio (SNR). 

To evaluate the viability of the system for spectroscopy applications under realistic conditions, we conducted a series of experiments involving a $\mathrm{CH_{4}}$ flame produced by a standard McKenna burner with a diameter of 7\,cm (see \textbf{Fig. \ref{fig:setup2}}). Positioned roughly 20\,cm from the dual-comb source and the detection optics, the burner was aligned to ensure optimal signal collection through the combustion region. The $\mathrm{CH_{4}}$ flow rate was modulated remotely, adjusting the equivalence ratio ($\gamma$) from fuel-lean ($\gamma = 0.7$) to fuel-rich ($\gamma = 1.5$), where $\gamma$ represents the $\mathrm{CH_{4}}$-to-air ratio relative to stoichiometric balance. 

By combining the EO comb generation and DFG with this dual-channel MIR detection scheme, the system achieves significant operational advantages. This architecture provides a high degree of flexibility in tuning comb tooth spacing and spectral spans, while maintaining the ruggedness and transportability necessary for practical field applications. In addition, the implementation of differential detection with a separate reference arm marks an improvement over prior designs, offering more accurate and reliable spectroscopic analysis of trace gases like $\mathrm{CH_{4}}$ in complex environments.

\subsection{Line Selection}

\begin{figure*}[ht!]
    \centering
    \begin{subfigure}[t]{0.48\textwidth}
    \begin{tikzpicture}[inner sep=0]
      \node[anchor=south west] (img) at (0,0)
        {\includegraphics[width=\textwidth]{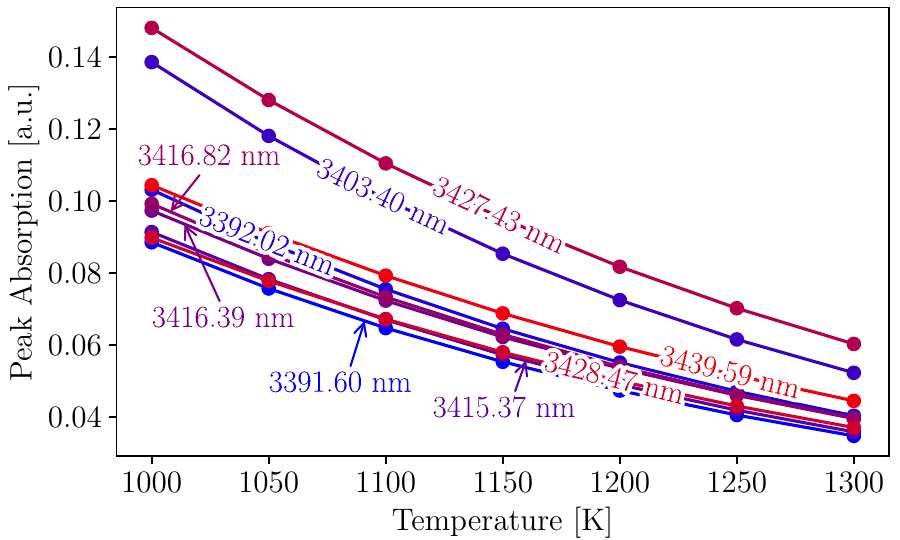}};
     \node[anchor=north west, xshift=-3pt, yshift=-3pt]
       at (img.north west) {\textbf{(a)}};
   \end{tikzpicture}
    \end{subfigure}
    \hfill
    \begin{subfigure}[t]{0.48\textwidth}
    \begin{tikzpicture}[inner sep=0]
      \node[anchor=south west] (img) at (0,0)
        {\includegraphics[width=\textwidth]{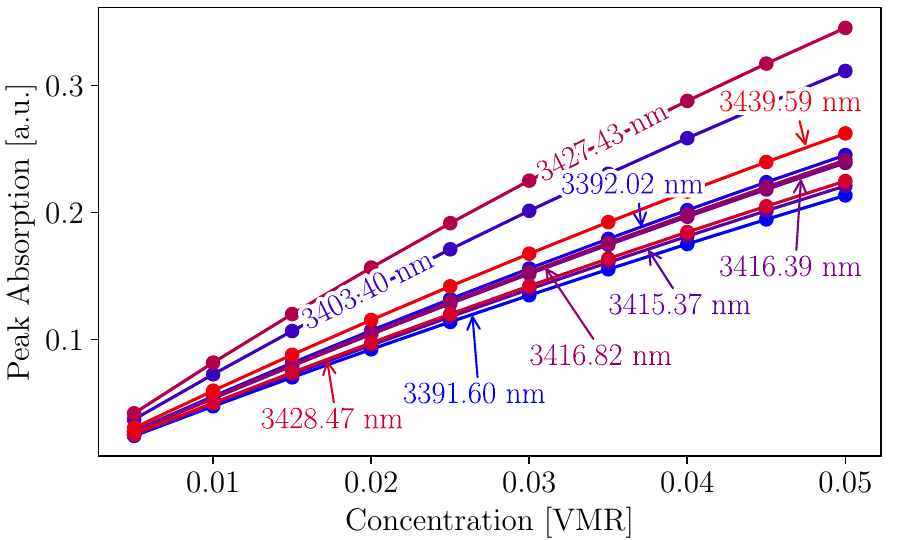}};
     \node[anchor=north west, xshift=-3pt, yshift=-3pt]
        at (img.north west) {\textbf{(b)}};
    \end{tikzpicture}
    \end{subfigure}
    \caption{Peak absorption for the $\mathrm{CH_{4}}$ spectral features between 3390\,nm and 3440\,nm with largest absorption as a function of (a) temperature for a concentration of 0.01\,VMR and (b) concentration for a temperature of 1200\,K. Simulations were done in both cases using HITEMP for a pressure of 1\,atm and a path length of 7\,cm.}
    \label{fig:peak_conc_temp}
\end{figure*}

The optimal $\mathrm{CH_{4}}$ absorption line was selected from the available spectral range of 3390-3440\,nm (limited by the DFG module) to maximize measurement performance. To begin, we studied the temperature dependence of the main spectral features in the range, and it was seen that all accessible lines exhibit significant temperature sensitivity within the expected operational range of 1000-1300\,K (see \textbf{Fig. \ref{fig:peak_conc_temp}a}). Consequently, it was determined that independent temperature measurements, obtained via a thermocouple, would be required as a fixed input for the fitting algorithm. With temperature being an independently measured parameter, the selection criteria were focused on signal fidelity and sensitivity based on two factors: (i) \textit{Maximum peak absorption} within the temperature range of interest, to achieve the highest possible SNR, and (ii) \textit{Maximum sensitivity to concentration}, identified by the steepest slope in the absorption versus concentration curve (\textbf{Fig. \ref{fig:peak_conc_temp}b}).

Based on these criteria, the $\mathrm{CH_{4}}$ absorption line at 3427.43\,nm was selected, as it offered the best combination of strong absorption and high sensitivity to concentration changes. Accordingly, the Toptica laser source was tuned to 1541.05\,nm and the Optilab pump laser operated at 1063.07\,nm.

\subsection{Comb configuration selection}

The architecture of our DC system permits the tuning of many comb parameters, such as the optical tooth spacing and the number of comb teeth ($N$), which are controlled by the modulation frequency and the input power (modulation intensity $\Omega$) of the PMs, respectively. However, a trade-off exists between spectral coverage, resolution, and SNR. A wider tooth spacing increases the spectral bandwidth at the cost of resolution. Similarly, while generating more comb teeth also broadens the bandwidth, it reduces the SNR of each tooth by distributing the available laser power over more optical teeth. Therefore, selection of these parameters must be done carefully to maximize the overall SNR across the targeted absorption feature, which is essential for achieving high-precision concentration measurements.

To determine the optimal comb configuration for measuring the 3427.43\,nm line, we conducted simulations using the $\mathrm{CH_4}$ HITEMP database \cite{Hargreaves_2020} via RADIS \cite{PANNIER201912}. By varying tooth spacing ($f_{r1}$) and number of comb teeth, we generated synthetic spectra that were subsequently fitted to derive concentration values. Simulation parameters of 0.01\,VMR (volume mixing ratio\footnote{The volume mixing ratio (VMR) is a number between 0 and 1 indicating the fractional number of molecules of a species in a volume. Individual VMRs must sum to 1.}), 1200\,K, 1\,atm, and a path length equivalent to the burner diameter were selected to approximate the experimental flame conditions. We ranked the configurations based on the fitting algorithm's accuracy in retrieving the true concentration. The parameter sweep included $f_{r1}$ values from 0.1\,GHz to 3\,GHz (increments of 100\,MHz) and $N$ values from 5 to 30.

Parameter combinations yielding excessively narrow or broad spectral widths were discarded. The analysis was restricted to comb bandwidths between 0.15\,nm, corresponding to the approximate full width at half maximum (FWHM) of the absorption line of interest, and 0.9\,nm, a width sufficient to capture the entire spectral feature and its surrounding baseline, leaving a total of 308 configurations.

The theoretical absorption line was obtained by sampling the HITEMP database at the specified comb frequencies. To model the output of a realistic measurement, these ideal profiles were altered to incorporate several non-ideal comb (most importantly noise) and instrumental (such as wavelength instability) effects which are described in \textbf{Supplementary Material \ref{appendixa}}.

\begin{figure*}[ht!]
    \centering
    \begin{subfigure}[t]{0.48\textwidth}
    \begin{tikzpicture}[inner sep=0]
      \node[anchor=south west] (img) at (0,0)
        {\includegraphics[width=\textwidth]{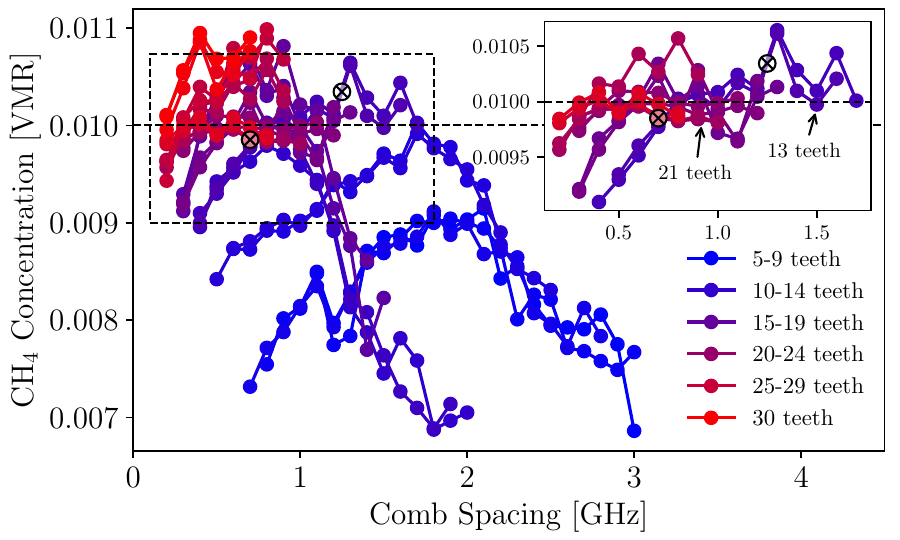}};
      \node[anchor=north west, xshift=-3pt, yshift=-3pt]
        at (img.north west) {\textbf{(a)}};
    \end{tikzpicture}
    \end{subfigure}
    \hfill
    \begin{subfigure}[t]{0.48\textwidth}
    \begin{tikzpicture}[inner sep=0]
      \node[anchor=south west] (img) at (0,0)
        {\includegraphics[width=\textwidth]{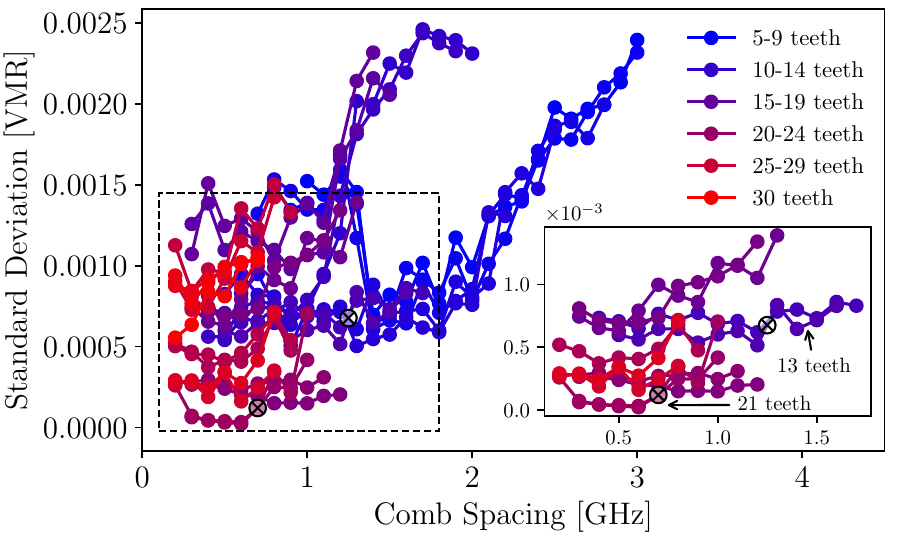}};
      \node[anchor=north west, xshift=-3pt, yshift=-3pt]
        at (img.north west) {\textbf{(b)}};
    \end{tikzpicture}
    \end{subfigure}
    \caption{Fitted $\mathrm{CH_4}$ concentration (a) and its standard deviation (b) on the simulated 0.01\,VMR measurements for the 308 studied comb configurations. Each point corresponds to the average or the standard deviation, respectively, of 100 simulated measurements and fittings. The insets show only the most viable comb configurations and the legend is given by ranges. The configurations that were finally selected are indicated by black ``x'' marks.}
    \label{fig:configs}
\end{figure*}

For each of the 308 configurations, 100 simulations were run, making a total of 30800 simulated measurements, and the mean and standard deviation of the fitted concentration for each configuration were recorded (see \textbf{Fig. \ref{fig:configs}a} and \textbf{Fig. \ref{fig:configs}b}). The simulations highlighted two key performance limitations, also later observed in experimental results: (i) \textit{underestimation with insufficient bandwidth} and (ii) \textit{increased variance with coarse spectral sampling}. The simulations show that configurations with a narrow spectral bandwidth systematically underestimate the concentration. This error arises because the comb fails to sample the true transmission baseline on the wings of the absorption feature, causing normalization to happen with respect to an artificially low baseline and yielding a reduced fitted absorbance. The increased variance is observed in configurations with large tooth spacing, as the absorption peak is poorly resolved, making the fit highly sensitive to noise or the exclusion of even just a single tooth. Experiments realized in the laboratory confirm that sparser combs are highly sensitive to the removal or exclusion of teeth near the absorption peak, while narrower combs fail to adequately resolve the spectral wings.

To isolate the most accurate results, configurations were discarded if their fitted concentration fell outside the $[0.009,0.011]$\,VMR interval. The retained configurations are presented in the insets of \textbf{Fig. \ref{fig:configs}a} and \textbf{Fig. \ref{fig:configs}b}.

The selection of an optimal configuration involves a trade-off between performance metrics such as accuracy and repeatability. This validation process singled out two setups. Configuration (i) used modulation frequencies of $f_{r1} = 1.25$\,GHz and $f_{r2} = 1.25$\,GHz $+$ 200\,Hz, yielding a 13-tooth comb with a bandwidth of 15\,GHz (0.59\,nm) that prioritizes high per-tooth power. Configuration (ii) used lower PM frequencies of 700\,MHz and 700\,MHz $+$ 200\,Hz, while maintaining the same AOM configuration, which produced a 21-tooth comb with a 14\,GHz (0.55\,nm) bandwidth chosen for its excellent predicted performance in simulations.

\section{Results and discussion}

To obtain the absorption measurements, two mid-infrared photodetectors were employed to record the dual-comb signal. Each photodetector down-converts the optical spectral information into the RF domain, yielding a spectral compression factor of $6.25 \times 10^6$. The resulting RF dual-comb features 13 or 21 (depending on the configuration used) comb lines with a mode spacing of $\Delta f=200$\,Hz. The outputs from both photodetectors, corresponding to the sample and reference branches, were simultaneously digitized at a sampling frequency of 400\,kHz over a one-second interval. The acquired time-domain data were then processed to extract concentration using a custom Python-based toolkit, which is publicly available via GitHub\footnote{\url{https://github.com/bfrangi/dual-comb-toolkit}} and archived on Zenodo \cite{bernat_frangi_2025_18040729}.

The analysis began by applying an FFT to the full-continuous time-domain data to obtain the RF spectra. The RF spectra were then mapped to the optical frequency domain, and the optical transmission spectrum was calculated by obtaining the ratio between the sample and reference spectra, compensating the baseline, and applying subsequent normalization. Besides, frequency components with excessive noise (with a standard deviation 1.5 times higher than the average) were discarded.

The $\mathrm{CH_{4}}$ concentration was then determined by fitting a theoretical spectrum, generated using the HITEMP database, to the measured optical spectrum, in the transmission domain. The fit was performed using the Nelder-Mead algorithm \cite{comjnl}, which minimized the mean absolute difference between the two spectra (simulated and measured). 

The theoretical sensitivity limit derived from the system's baseline noise was determined by recording a 1\,s baseline spectrum with an empty optical path. This measurement was segmented into ten 100\,ms intervals to calculate the amplitude standard deviation per comb tooth. After normalizing to a 1\,s integration time (scaling by $\sqrt{10}$) and averaging across the spectrum, we obtained a mean standard deviation of $9.55\times 10^{-4}$, which corresponds to a detection limit of 1.1\,ppm for a 1\,m path length and a 1\,s integration time. To substantiate this sensitivity and demonstrate the system's stability, an Allan deviation analysis was performed (shown in the inset (i) of \textbf{Fig. \ref{fig:time-evolution}}). The plot confirms an ideal scaling of the noise over time, verifying that the integration times used for our measurements fall well within the system's coherence limits.

To validate the one-second integration time regarding flame stability, we conducted 5-second measurements at $z=1$\,mm and $z=22$\,mm  (maximum spatial range provided by the translation stage) above the burner base for $\gamma$ values ranging from 0.7 to 1.5. These data were segmented into 100, 50, and 25\,ms intervals to illustrate the time-resolved concentration evolution (example in \textbf{Fig. \ref{fig:time-evolution}}). Subsequent FFT analysis consistently identified two primary oscillatory components of comparable amplitude: a low-frequency mode near 1\,Hz and a higher-frequency oscillation at approximately 8--9\,Hz (inset (ii) of \textbf{Fig. \ref{fig:time-evolution}}). The detection limit for the 25\,ms integration time used in \textbf{Fig. \ref{fig:time-evolution}} is estimated at approximately 99\,ppm within the 7\,cm optical path length of the burner, which is an order of magnitude lower than the observed oscillation amplitudes. This confirms that these variations in methane concentration are flame instabilities rather than a measurement artifact.

\begin{Figure}
    \centering
    \includegraphics[width=1\textwidth]{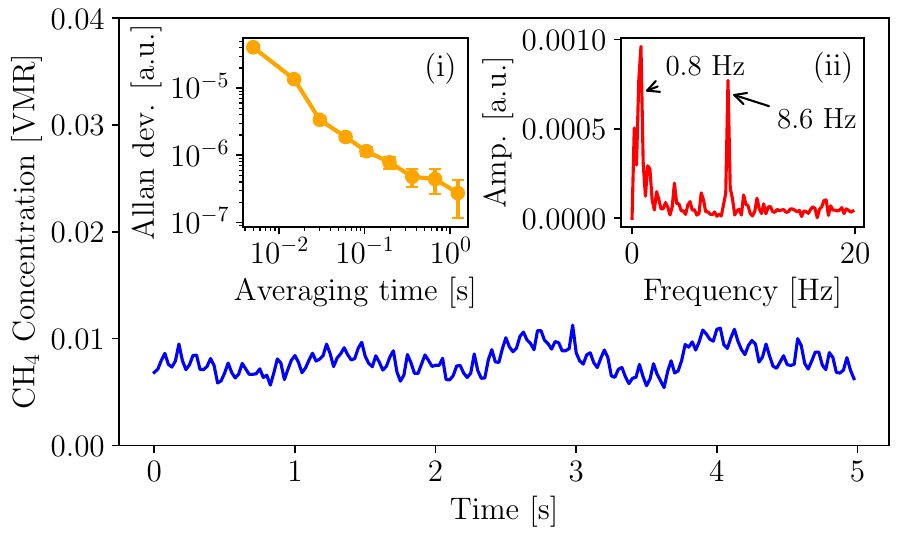}
    \captionof{figure}{Evolution of $\mathrm{CH_{4}}$ concentration in a 5-second continuous measurement at the burner base ($z=1$\,mm) for an equivalence ratio of $\gamma = 0.7$. The chosen 13-tooth configuration was used and the complete measurement was analyzed in small intervals of 0.025\,s. Inset (i) shows the Allan deviation plot demonstrating the ideal noise scaling of the system over the relevant integration times. Inset (ii) shows the FFT of the concentration evolution, where the zero-frequency component has been removed.}
    \label{fig:time-evolution}
\end{Figure}
\begin{figure*}[hb!]
    \centering
    \begin{subfigure}[t]{0.48\textwidth}
    \begin{tikzpicture}[inner sep=0]
      \node[anchor=south west] (img) at (0,0)
        {\includegraphics[width=\textwidth]{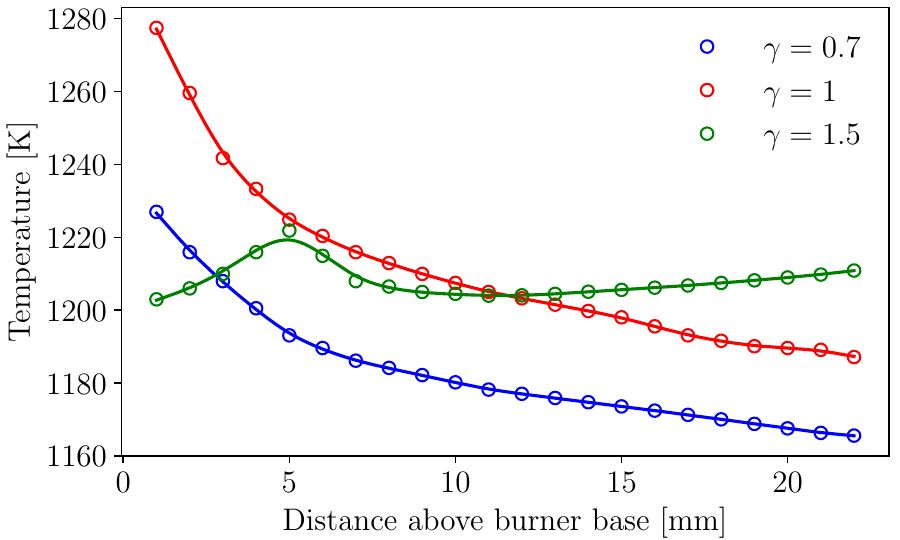}};
      \node[anchor=north west, xshift=-3pt, yshift=-3pt]
        at (img.north west) {\textbf{(a)}};
    \end{tikzpicture}
    \end{subfigure}
    \hfill
    \begin{subfigure}[t]{0.48\textwidth}
    \begin{tikzpicture}[inner sep=0]
      \node[anchor=south west] (img) at (0,0)
        {\includegraphics[width=\textwidth]{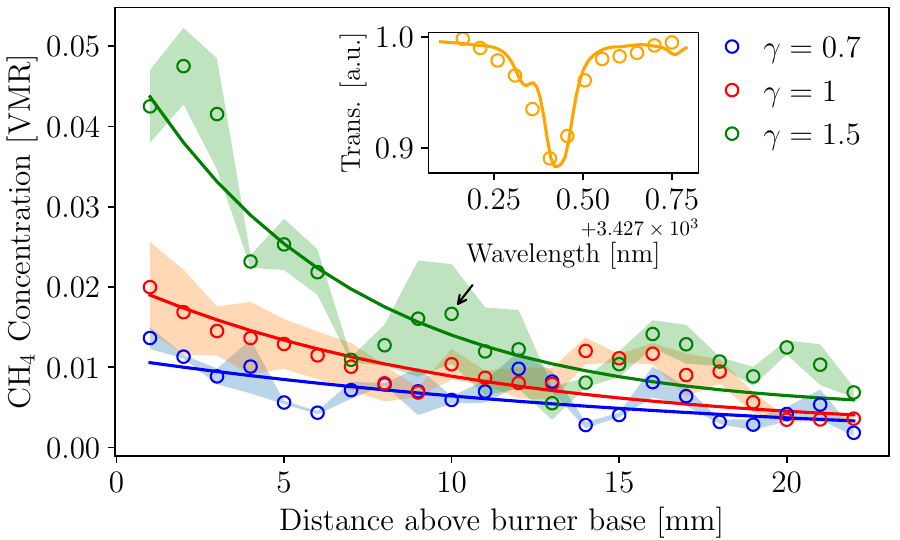}};
      \node[anchor=north west, xshift=-3pt, yshift=-3pt]
        at (img.north west) {\textbf{(b)}};
    \end{tikzpicture}
    \end{subfigure}
    \caption{Measured temperature (a) and $\mathrm{CH_4}$ concentration (b) as a function of distance above the burner base for different values of $\gamma$. Temperature was measured using a thermocouple and used as an input to the fitting algorithm. Concentration is an average of $3$ measurements, each obtained using either the $13$-tooth or the $21$-tooth comb configuration. Shaded areas represent the standard deviation at each spatial point. This variance is primarily driven by dynamic flame fluctuations---specifically, slow oscillations at approximately 0.8 Hz---which dominate the measurement uncertainty. Solid lines represent the trend. The inset shows a typical measured spectrum along with the fitted HITEMP spectrum.}
    \label{fig:concs-temps}
\end{figure*}

The stabilization of flat flames over a porous plug burner is known to exhibit unstable characteristics under certain operating conditions \cite{MISLAVSKII2021111638, nie2025pulsationburnerstabilizedch4o2flames}. Specifically, when the equivalence ratio is fixed, a decrease in the mass flow rate causes the flame to stabilize closer to the burner surface. This smaller standoff distance enhances heat losses, leading to a significant reduction in the burned gas temperature. This thermal effect causes the initially steady, planar flame front to lose stability, inducing a self-sustained pulsating regime. The frequency of the resulting oscillation is primarily governed by the equivalence ratio and the incoming flow rate.

Our experiments confirm this instability and further reveal significant oscillations in the downstream $\mathrm{CH_{4}}$ concentration, even for fuel-lean mixtures ($\gamma=0.7$), as shown in \textbf{Fig. \ref{fig:time-evolution}}. This $\mathrm{CH_{4}}$ leakage takes place near the burner's perimeter, where buoyancy effects cause the flame to lift and curve. This phenomenon opens a cooler region at the base that provides a passage for unburned $\mathrm{CH_{4}}$ to escape.

Once this validation was performed, the measurement and fitting procedure was repeated for 22 axial positions above the burner, with a spatial resolution of 1\,mm. The complete scanning process required less than one minute, with a large part of the time taken up by communication latency between the control software and the translation stage actuators as well as their movement. 

The spatial distribution of unburned $\mathrm{CH_{4}}$ concentration was profiled for three fuel-air equivalence ratios: fuel-lean ($\gamma=0.7$), stoichiometric ($\gamma=1.0$), and fuel-rich ($\gamma=1.5$). To ensure reproducibility and reduce uncertainty, this spatial scan was repeated three times for each $\gamma$. Each scan was performed using either the 13- or 21-tooth configuration and processed independently, yielding three concentration profiles per equivalence ratio. The final data points plotted in \textbf{Fig. \ref{fig:concs-temps}b} represent the average of these three independent concentration profiles for each $\gamma$. While both setups yielded highly consistent and similar results, the 13-tooth comb provided slightly superior performance in practice, as its power distribution aligned more closely with the theoretical Bessel model. During processing of each spatial scan, individual comb lines exhibiting a low SNR were excluded from the analysis prior to the fitting procedure. The theoretical spectrum was simulated at a fixed pressure of 1 atm using the temperature profiles measured for each value of $\gamma$ (see \textbf{Fig. \ref{fig:concs-temps}a}). Some studies have estimated the uncertainty in this approach to the measurement of temperature to be 7\% \cite{10.1063/5.0176359}. Hence, future work will explore a simultaneous fit for both temperature and concentration to improve accuracy. The distinct temperature trend observed for $\gamma=1.5$ is a direct consequence of the flame's structure, which at rich methane content features both a premixed flame near the burner surface and a diffusion flame further downstream. \textbf{Fig. \ref{fig:concs-temps}b} reports the average concentration of these repeated measurements.

Error analysis of the spatial profiles indicates that instrumental noise is negligible compared to the inherent physical fluctuations of the flame. The observed variance is primarily driven by slow, macroscopic flame oscillations at approximately 0.8 Hz. To represent this physical variability, the shaded regions in \textbf{Fig. \ref{fig:concs-temps}b} indicate the standard deviation from the mean of the three independent measurements, capturing the dynamic nature of the flame rather than instrumental uncertainty. Despite these physical fluctuations, the system's ability to clearly distinguish overarching spatial trends across varying equivalence ratios remains evident.

In light of these results, it is useful to compare the presented system with prior dual-comb implementations. Compared to conventional NIR systems based on mode-locked lasers \cite{PhysRevLett.100.013902, Zolot:12}, the architecture here proposed features a much simpler implementation while still providing high performance, enabling a single-pass detection limit of 1.1\,ppm in the detection of $\mathrm{CH_4}$ for a 1\,s integration time. Additionally, the use of an optical fiber allows for an easy integration of the interrogation optics onto a translation stage, allowing us to perform millimeter-scale spatially resolved measurements in open and dynamic combustion environments.

Even compared to alternative MIR systems such as QCLs \cite{Hugi2012} and Kerr frequency combs \cite{Wang2013}, which distribute power over hundreds of thousands of teeth, our EO approach concentrates power into a small number of teeth tailored to the absorption feature. This results in significantly higher power per tooth and improved SNR, enabling fast measurements and the capture of dynamic flame behavior.

Finally, the use of a reference optical channel provides a key advantage for combustion diagnostics, compensating both the comb power envelope and fluctuations across individual teeth, and enabling precise, calibration-free spectral measurements without complex baseline correction.

\section{Conclusions}

This study successfully reports the development and implementation of a robust mid-infrared DC spectroscopic system utilizing EO frequency comb generators for the characterization of laboratory flames. By employing a differential detection strategy, the system enabled precise, calibration-free spectral measurements of unburned $\mathrm{CH_{4}}$ concentrations in a McKenna burner, achieving a detection limit of 1.1 ppm for a 1\,m path length with a 1\,s integration time.

The experimental results demonstrated the system's capacity to resolve spatial concentration profiles and identify dynamic combustion instabilities, specifically revealing oscillatory behavior and fuel leakage near the burner perimeter. Future developments may include the implementation of cavity-enhanced architectures to extend the effective path length \cite{SUN2024105662}, as well as the exploration of quantum-correlation based dual-comb schemes capable of surpassing the shot noise limit \cite{Wan2025}, with the aim of achieving faster and ultra-sensitive combustion diagnostics. Ultimately, this work establishes the viability of EO DC spectroscopy as a flexible and high-sensitivity tool for advanced combustion diagnostics in complex environments.

\section{CRediT authorship contribution statement}

\textbf{Bernat Frangi}: 
Data curation,
Formal analysis,
Investigation,
Software,
Visualization,
Writing – original draft,
Writing – review \& editing.
\textbf{Laura Monroy}:
Investigation,
Methodology,
Software,
Validation,
Visualization,
Writing – original draft,
Writing – review \& editing.
\textbf{Aldo Moreno-Oyervides}:
Methodology,
Validation,
Writing – review \& editing.
\textbf{Oscar El\'ias Bonilla-Manrique}:
Methodology,
Validation.
\textbf{Mariano Rubio-Rubio}:
Investigation,
Methodology,
Resources,
Validation,
Writing – review \& editing.
\textbf{Mario S\'anchez-Sanz}:
Conceptualization,
Project administration,
Resources,
Supervision,
Writing – review \& editing.
\textbf{Pedro Mart\'in-Mateos}:
Conceptualization,
Funding acquisition,
Investigation,
Methodology,
Project administration,
Supervision,
Validation,
Writing – original draft,
Writing – review \& editing.

\section{Declaration of Competing Interest}
The authors declare that they have no known competing financial interests or personal relationships that could have appeared to influence the work reported in this paper.

\section{Acknowledgments}
This work has been funded by the R\&D programme with reference TEC-2024/ECO-99 and acronym HyCoTec-CM granted by the Community of Madrid through the Directorate-General for Research and Technological Innovation by Order 5696/2024, of 10 December, of the Regional Minister of Education, Science and Universities.

\section{Data availability}
Data will be made available on request.

\bibliography{sample}

\appendix
\numberwithin{equation}{section}
\numberwithin{figure}{section}

\newcommand{\appsection}[1]{%
\refstepcounter{section}%
\section*{Appendix \thesection: #1}%
\addcontentsline{toc}{section}{Appendix \thesection: #1}%
}

\appsection{Supplementary material}\label{appendixa}

\subsection{Non-ideal comb considerations}

\begin{figure*}[htb]
    \centering
    \begin{subfigure}[t]{0.48\textwidth}
    \begin{tikzpicture}[inner sep=0]
      \node[anchor=south west] (img) at (0,0)
        {\includegraphics[width=\textwidth]{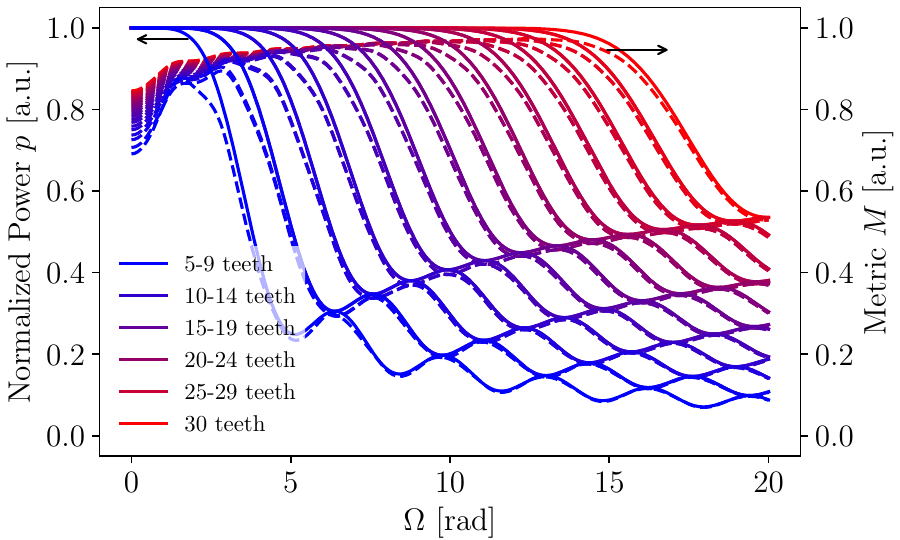}};
      \node[anchor=north west, xshift=-3pt, yshift=-3pt]
        at (img.north west) {\textbf{(a)}};
    \end{tikzpicture}
    \end{subfigure}
    \hfill
    \begin{subfigure}[t]{0.48\textwidth}
    \begin{tikzpicture}[inner sep=0]
      \node[anchor=south west] (img) at (0,0)
        {\includegraphics[width=\textwidth]{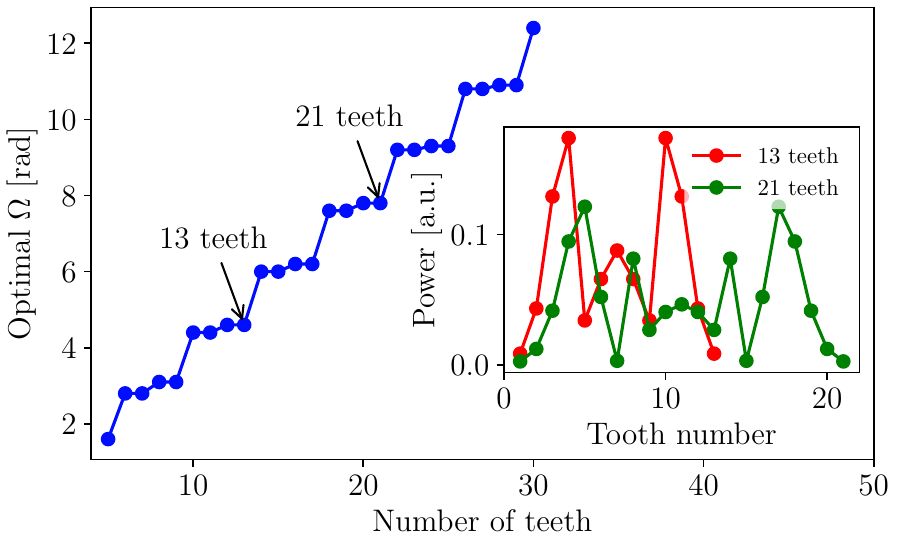}};
      \node[anchor=north west, xshift=-3pt, yshift=-3pt]
        at (img.north west) {\textbf{(b)}};
    \end{tikzpicture}
    \end{subfigure}
    \caption{(a) Normalized total power (solid lines) and optimality metric (dashed lines) for different values of $N$, the number of comb teeth. The legend is given by ranges. (b) Optimal modulation intensity $\Omega$ for the different values of $N$, the number of comb teeth. The optimization favors a step-like distribution of the optimal $\Omega$ to avoid quasi-periodically spaced zeros of the Bessel functions, which cause complete loss of power for some teeth at regularly spaced values of $\Omega$. The inset shows the power distribution among the teeth for the chosen configurations of $13$ and $21$ teeth.}
    \label{fig:omega-selection}
\end{figure*}

In phase modulation, the phase of a carrier signal $E_0e^{i\omega_c t}$ is modulated by an external voltage source of frequency $\omega$:
\begin{equation}
    E_0 \mathrm{e}^{i\omega_c t + i\Omega \sin(\omega t) }.
\end{equation}
Using the Jacobi-Anger expansion \cite{schenzle1982phase}, this becomes:

\begin{equation}
    E_0\mathrm{e}^{i \omega_c t} \biggl(J_0(\Omega) + \sum_{k=1}^{\infty} J_k(\Omega) \mathrm{e}^{i k \omega t}
 +\sum_{k=1}^{\infty} (-1)^k J_k(\Omega) \mathrm{e}^{-i k \omega t}\biggl),  
\end{equation}
    
where $J_k(\Omega)$ are the $k$-th order Bessel functions of the first kind evaluated at the intensity of the modulation $\Omega$. The result is an infinite number of sidebands appearing at either side of the carrier frequency $\omega_c$, which form the teeth of the comb. The power of the $k$-th sideband is proportional to $\mid J_k(\Omega)\mid ^2$, which decays for large $k$ at rates dependent on $\Omega$, thereby limiting the usable bandwidth for a given modulation intensity. Moreover, the power distribution among comb teeth within this bandwidth is nonuniform and also varies with $\Omega$, so not all teeth contribute equally to the noise.

In the simulations, the optimal $\Omega$ for each number of teeth was determined by means of an optimality metric that takes into account the tradeoff between (i) normalized comb power $p(\Omega)$, obtained as the sum of the $|J_k(\Omega)|^2$ terms for each tooth; and (ii) uniformity of the power distribution among comb teeth, measured by the standard deviation $\sigma(\Omega)$ of the $|J_k(\Omega)|^2$ terms associated with comb teeth. The optimality metric is a dimensionless number defined as:
\begin{equation}
    M(\Omega) = \frac{p (\Omega)}{1 + \sigma(\Omega)}.
\end{equation}

\balance{
The total comb power and the optimality metric have been computed for values of $\Omega$ between $0$ and $20$ rad for the different values of $N$, the number of comb teeth (see \textbf{Fig. \ref{fig:omega-selection}a}), and the modulation intensities with highest $M(\Omega)$ have been selected (see \textbf{Fig. \ref{fig:omega-selection}b}).

To simulate the comb SNR, Gaussian noise was added to the comb teeth. For the $k$-th tooth having an SNR of $1/\sigma_k$, being $\sigma_k$ the standard deviation, a Gaussian distribution of zero mean was used to simulate its noise. The standard deviation $\sigma_k$ was obtained in the following manner:
\begin{enumerate}
    \item A measurement of the $3427.43$ nm absorption line at room temperature was taken using a $30$-teeth dual comb.
    \item The same line was simulated at the same conditions using the HITRAN database \cite{GORDON2022107949}, appropriate for room-temperature measurements.
    \item The average standard deviation between the measurement and the HITRAN simulation was found to be $0.014$.
    \item For a simulation of $N$ teeth, the average standard deviation was scaled according to \cite{Coddington:16}:
    \begin{equation}
        \overline \sigma(N) = 0.014\cdot \frac{N}{30} = \frac{N}{2143}.
    \end{equation}
    \item The values of $1/\mid J_k(\Omega)\mid ^2$ were obtained for each tooth, where $J_k(\Omega)$ is the $k$-th Bessel function evaluated at the modulation intensity $\Omega$, and the average of all teeth, $\overline{1/J^2}$, was computed.
    \item The standard deviation $\sigma_k$ of the $k$-th tooth was obtained as:
    \begin{equation}
        \sigma_k = \overline{ \sigma}\cdot \frac{1/\mid J_k ( \Omega)\mid^2}{\overline{1/J^2}}.
    \end{equation}
\end{enumerate}

The following additional considerations were made to simulate instrumental effects:

\textit{Laser wavelength instability}. Two sequential effects were modeled to account for instabilities in the laser's central wavelength.

\begin{enumerate}[label=(\roman*)]
    \item Shot-to-shot jitter: A random frequency offset was applied to the comb's entire optical frequency grid \textit{prior} to sampling the HITEMP line-shape. This simulates the physical scenario where the comb's sampling points shift relative to the fixed absorption feature between measurements. Consequently, each tooth probes a slightly different point along the absorption curve in successive acquisitions. This offset was drawn from a uniform distribution over the interval $[-f_{r1}, +f_{r1}]$, where $f_{r1}$ is the optical tooth spacing.
    \item Wavelength mapping error: A systematic error was modeled to account for the discrepancy between the true central wavelength and the nominal value used in post-processing to map the RF spectrum to the optical domain. This was implemented by applying an additional random wavelength shift to the entire simulated spectrum after it was generated from the HITEMP data. This shift was drawn from a uniform distribution over $[-0.02, +0.02]$ nm.
\end{enumerate}

\textit{Baseline amplitude variation}. Due to unbalanced power splitting between the sample and reference paths, the measured transmission baseline deviates from unity. To account for this, each simulated spectrum was multiplied by a random scaling factor drawn from a uniform distribution over the interval $[0.2, 1.5]$, thereby testing the robustness of the algorithm's normalization procedure.

\textit{Adaptive noise-based filtering}. An adaptive outlier rejection filter was applied, removing any comb tooth $k$ whose amplitude standard deviation $\sigma_k$ exceeded the mean standard deviation $\overline{\sigma}$ by a variable factor $t_\sigma$ (i.e., $\sigma_k > t_\sigma \cdot \overline{\sigma}$). The factor $t_\sigma$ was decreased linearly from $2.5$ to $1.5$ as the number of teeth $N$ increased from $5$ to $30$. This less stringent threshold for sparser combs prevents the detrimental removal of critical data points. The specific values were determined empirically to yield the most robust fitting results.}

\end{multicols}
\end{document}